\documentclass[11pt]{article}
\RequirePackage{amsgen}[1995/01/01]
\RequirePackage{amsfonts}[1995/01/01]
\RequirePackage{amssymb}[1995/01/01]
\RequirePackage{amsbsy}[1995/01/01]
\usepackage[margin=0.5in]{geometry}
\usepackage[dvips]{graphicx}
\usepackage[english]{babel}
\usepackage{verbatim}
\usepackage{amsfonts}
\usepackage{makeidx}
\usepackage{color}
\usepackage{amsmath}
\usepackage{subfigure}
\usepackage{multirow}
\usepackage{setspace}
\newcommand{\pp}{\rm pp}
\newcommand{\sqrts}{\sqrt{s}}
\newcommand{\sqrtsNN}{\sqrt{s_{\rm NN}}}
\newcommand{\gsim}{\,{\buildrel > \over {_\sim}}\,}

\newcommand{\ctau}{{\rm c}\tau}

\newcommand{\GeV}{\mathrm{GeV}}
\newcommand{\TeV}{\mathrm{TeV}}

\newcommand{\gev}{\mathrm{GeV}}
\newcommand{\tev}{\mathrm{TeV}}

\newcommand{\cm}{\mathrm{cm}}

\newcommand{\mum}{\mathrm{\mu m}}

\newcommand{\mub}{\mathrm{\mu b}}

\newcommand{\ptrans}{p_{\rm t}}

\newcommand{\DtoKpi}{{\rm D^0\to K^-\pi^+}}
\newcommand{\DtoKpipi}{{\rm D^+\to K^-\pi^+\pi^+}}
\newcommand{\DstartoDpi}{{\rm D^{*+}\to D^0\pi^+}}
\newcommand{\Dzero}{{\rm D^0}}
\newcommand{\Dstar}{{\rm D^{*+}}}
\newcommand{\Dplus}{{\rm D^+}}

\newcommand{\vtwo}{v_{2}}
\newcommand{\RAA}{R_{\rm AA}}
\newcommand{\RAAD}{R_{\rm AA}^{\rm D}}
\newcommand{\RAAB}{R_{\rm AA}^{\rm B}}

\newcommand{\Jpsi}{{\rm J/}\psi}

\newcommand{\avNcoll}{\langle N_{\rm coll} \rangle }
\newcommand{\avTAA}{\langle T_{\rm AA} \rangle}

\newcommand{\Npart}{\langle N_{\rm part} \rangle}

\title{Open heavy-flavour production in pp and Pb--Pb collisions at the LHC, measured with ALICE at central rapidity}
\author{A~Rossi for the ALICE Collaboration\\
Universit\`a di Padova and INFN - Sezione di Padova, Padova, Italy - now at CERN \\
email: andrea.rossi@cern.ch
}
\begin{document}
\maketitle

\begin{abstract}
The ALICE experiment studies nucleus–-nucleus collisions at the LHC in order to investigate 
the properties of QCD matter at extreme energy densities. 
The measurement of open charm and open beauty production allows to investigate the 
interaction of heavy quarks with the hot and dense medium formed in high-energy nucleus-nucleus 
collisions. In particular, in-medium energy loss is predicted to be different for 
gluons, light quarks and heavy quarks and to depend on the medium energy density and size.  
ALICE has measured open heavy-flavour particle production at central rapidity
in several decay channels in Pb-Pb and pp collisions at $\sqrtsNN = 2.76~\TeV$ and $\sqrts = 2.76$, $7~\TeV$ respectively. 
The results obtained from the reconstruction of D meson decays at central rapidity and 
from electrons from heavy-flavour hadron decay will be presented. 

\end{abstract}
%
\section{Introduction}
The comparison of heavy-flavour hadron production in proton-proton and Pb--Pb collisions
at the LHC offers the opportunity to investigate the properties of the high-density
colour-deconfined state of strongly-interacting matter (Quark Gluon Plasma, QGP) that is expected to be formed
in high-energy collisions of heavy nuclei. 
Due to their large mass, charm and beauty quarks are 
created in hard-scattering processes with high virtuality ($Q^{2}\gsim 4m_{c[b]}^{2}$)
involving partons of the incident nuclei, hence at the initial stage of the collision.
They interact with the medium and lose energy 
via both inelastic (medium-induced gluon radiation, or radiative energy loss)~\cite{gyulassy,bdmps} 
and elastic (collisional energy loss)~\cite{thoma} processes. This in-medium energy loss, which 
is sensitive to the medium energy density, is expected to affect 
differently heavy quarks, gluons and light quarks.   
In QCD, quarks have a smaller colour coupling factor with respect to gluons, 
so that the energy loss for quarks is expected to be smaller than for gluons. 
In addition, the `dead-cone effect' should reduce small-angle gluon radiation for heavy quarks with moderate 
energy-over-mass values~\cite{deadcone}. 
A sensitive observable is the nuclear modification factor,
defined, for a particle species $h$, 
as $\RAA^{h}(\ptrans)=\frac{{\rm d}N^{h}_{\rm AA}/{\rm d}\ptrans}{\avTAA\times {\rm d}\sigma^{h}_{\pp}/{\rm d}\ptrans}$,
where $N^{h}_{\rm AA}$ is the 
yield measured
in heavy-ion collisions, $\avTAA$ is the average nuclear overlap function calculated with the Glauber model~\cite{glauber} in the
considered centrality range, and $\sigma^{h}_{\pp}$ is the production cross section 
in pp collisions. 
In-medium energy loss determines a suppression, $\RAA<1$, of hadrons at 
moderate-to-high transverse momentum ($\ptrans\gsim 2~\gev/c$).
By comparing the nuclear modification factors 
of pions $(\RAA^{\pi})$, mostly originating
from gluon fragmentation, with that of hadrons with charm $(\RAAD)$ and beauty $(\RAAB)$
the dependence of the energy loss on the parton nature (quark/gluon) and mass can be investigated.
A mass ordering pattern $\RAA^{\pi}(\ptrans)\lesssim \RAAD(\ptrans)\lesssim \RAAB(\ptrans)$ has been predicted~\cite{deadcone,Armesto:2005iq}.
ALICE has studied open heavy-flavour production in pp and Pb--Pb collisions at mid-rapidity by measuring 
the $\ptrans$-differential production cross sections of electrons from heavy-flavour hadron decays~\cite{Masciocchi2011,BailhacheSQM} 
and of $\Dzero$, $\Dplus$ and $\Dstar$ mesons via exclusive reconstruction of hadronic decay channels~\cite{Dpp7TeVpaper,RAADmeson}. 
At forward rapidity, ALICE has measured the production of single muons coming from heavy-flavour hadron decays~\cite{LoicPrimQCD}. 
The large suppression observed in Pb--Pb collisions w.r.t. pp collisions, 
about a factor 3-4 for $\ptrans\gsim 5~\GeV/c$, suggests that charm quarks 
strongly interact with the medium. It is natural then to wonder whether charm is thermalized.\\

In heavy-ion collisions with non-zero impact parameter the interaction region exhibits 
an azimuthal asymmetry with respect to the reaction plane ($\Psi_{\rm RP}$), defined
by the azimuth of the impact parameter and the beam direction. Collective effects
convert this geometrical anisotropy into an asymmetry in momentum space that
is reflected in the final hadron azimuthal distribution~\cite{ArmestoFlow}.
The effect, sensitive to the degree of thermalization of the system,
can be evaluated by measuring the $2^{\rm nd}$ coefficient 
of the Fourier expansion of the azimuthal distribution, called elliptic
flow ($\vtwo$)~\cite{ArmestoFlow}. The measurement of the heavy-flavour particle $\vtwo$ 
can provide fundamental information on the degree of thermalization 
of charm quarks in the medium.
Some models~\cite{MolnarFlow} predict charm elliptic flow to be smaller than that of
light quarks at $\ptrans$ up to $\sim 2~\GeV/c$ and comparable at higher $\ptrans$. 
The PHENIX experiment at RHIC has measured 
the heavy-quark flow via non-photonic electrons, observing non-zero values 
($\vtwo\sim 0.09$ at $\ptrans\sim 1.5~\GeV/c$)~\cite{PhenixFlow}.
ALICE has performed a preliminary measurement of the $\Dzero$ elliptic
flow~\cite{BianchinFlow}.

In these proceedings the measurements of the nuclear modification factor
of electrons from heavy-flavour hadron 
decays $\RAA$ and of $\Dzero$,$\Dplus$ and $\Dstar$ are reviewed
as well as the preliminary measurement of the $\Dzero$ $\vtwo$.
%
%
%
\section{ALICE heavy-flavour probes at central rapidity}
\begin{figure}[!t]
  \begin{center}
    \includegraphics[height=0.3\textheight]{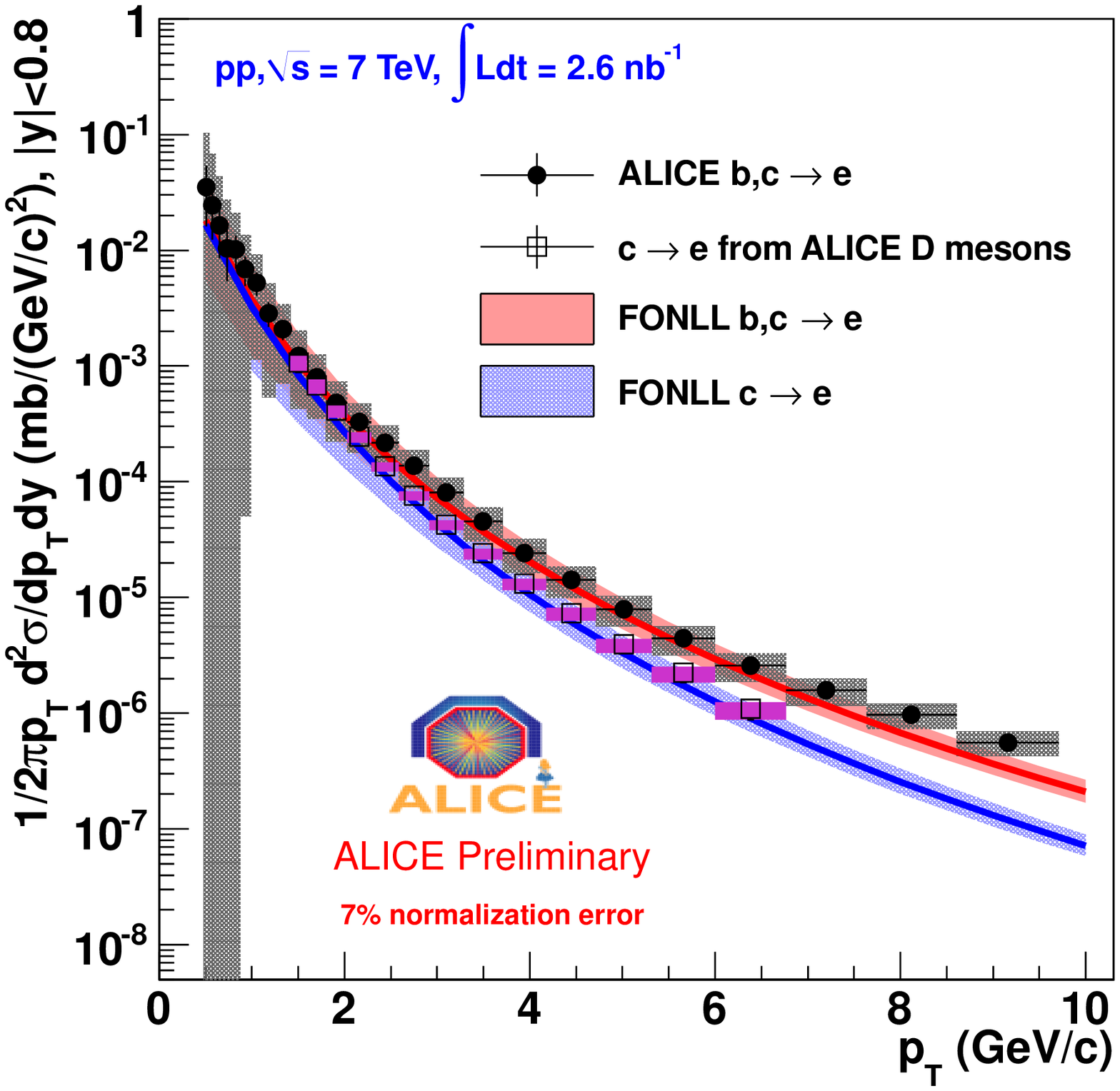}
    \includegraphics[height=0.28\textheight]{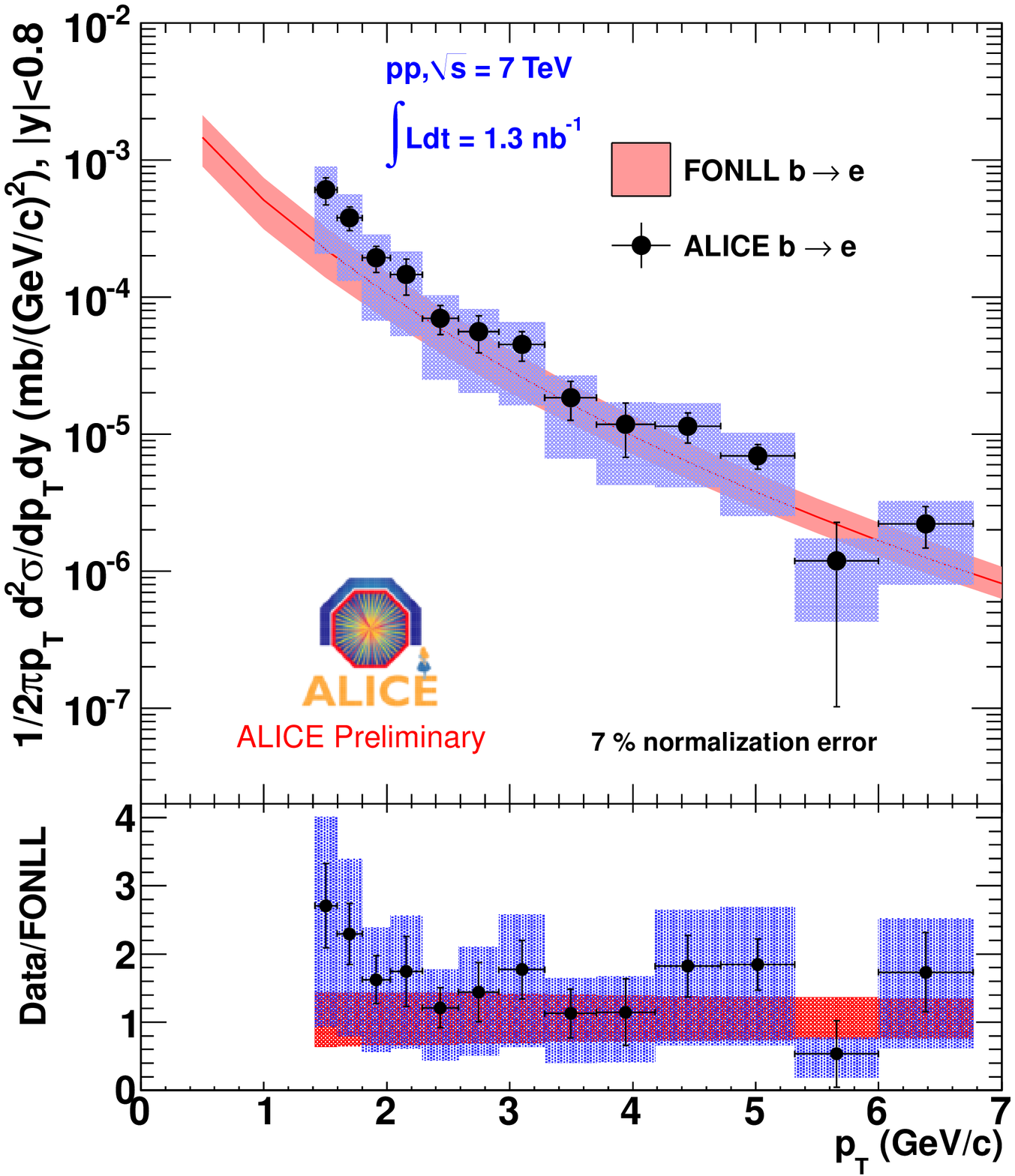}
    \caption[]{$\ptrans$-differential production cross section of electrons from the decay of hadrons 
      carrying a charm or beauty quark (on the left), or a beauty quark (on the right) in pp collisions
      at $\sqrts=7~\TeV$. The
      measurements are compared to FONLL predictions~\cite{fonll} (see text).
    }
    \label{fig:electronpp}    
  \end{center}
\end{figure}
The ALICE detector, described in detail in~\cite{aliceJINST}, consists of 
a central barrel composed of various detectors for particle reconstruction 
at midrapidity, a forward muon spectrometer, 
and a set of forward detectors for triggering and event characterization.
In particular, in the central pseudo-rapidity 
region ($|\eta|<0.9$), tracks are reconstructed
in the Time Projection Chamber (TPC) and in the Inner Tracking System (ITS),
and then propagated out to the Transition Radiation Detector (TRD) 
and to the the Time Of Flight (TOF) detector. 
An Electromagnetic Calorimeter (EMCal) 
covers the pseudo-rapidity range $|\eta|<0.7$ with an azimuthal acceptance
of $\Delta\phi=107^{\circ}$, limited to $\Delta\phi=40^{\circ}$ in 2010 when the detector
was not fully installed.
The mentioned detectors are embedded in a 0.5~T magnetic field parallel to the LHC beam 
direction ($z$-axis in the ALICE reference frame).
The VZERO detector, composed of two arrays of scintillator tiles covering
the full azimuth in the pseudo-rapidity regions $2.8 < \eta < 5.1$ (VZERO-A)
and $-3.7 < \eta < -1.7$ (VZERO-C), is used for triggering
and for the determination of the event centrality, as described in detail in~\cite{AlbericaProcQM}. 
\subsection{Electrons from heavy-flavour hadron decay}
Electrons are identified at mid-rapidity ($|{\eta}|<0.8$) using the information
provided by the TPC, the TOF and the TRD. The EMCal is being included as well. 
The remaining hadron contamination is determined via fits of the TPC $\rm{d}E/{\rm d}x$
in momentum slices and subtracted from the electron spectra.  More details on the adopted electron identification criteria
can be found in~\cite{Masciocchi2011,BailhacheSQM}.
In pp collisions, electrons are measured from $\ptrans=0.5~\GeV/c$ to $10~\GeV/c$. The 
hadron contamination amounts to less than $5\%$. In Pb--Pb collisions,
electrons are measured from $1.5~\GeV/c$ to $6~\GeV/c$. The TRD has not been 
included yet in the analysis. The hadron contamination is therefore larger than in the pp case and reaches
about $10\%$ at $6~\GeV/c$.

The inclusive electron yield is corrected for acceptance,
tracking and particle identification efficiency using Monte Carlo simulation.
In addition to electrons
from heavy-flavour hadron decays, that constitute the main contribution
at high $\ptrans$ ($\ptrans \gsim 2~\GeV/c$ in pp collisions),
the inclusive spectrum contains background electrons from 
Dalitz decay of light mesons (mainly $\pi^{0}$),
from the conversion in the material of photons from $\pi^{0}$ decay, Dalitz decays and direct 
radiation, and from dielectron decays of vector mesons (mainly from $\rho$,$\omega$,$\phi$,
at low $\ptrans$, also from heavy quarkonia at high $\ptrans$).
Photon conversion and $\pi^{0}$ Dalitz decay represent
the largest background contribution. To reduce the latter contribution 
the tracks used in the analysis are required to have a hit in the innermost
pixel layer, positioned at a radius of $3.9~\cm$ from the beam line.
In order to obtain the spectrum of electrons from heavy-flavour hadron decay,
a MC event generator is used to calculate a ``cocktail'' 
of background electrons, starting from the pion  
spectrum measured with ALICE and relying on $m_{\rm t}$-scaling
for the heavier mesons~\cite{Masciocchi2011}.
\begin{figure}[!t]
  \begin{center}
    \includegraphics[height=0.25\textheight]{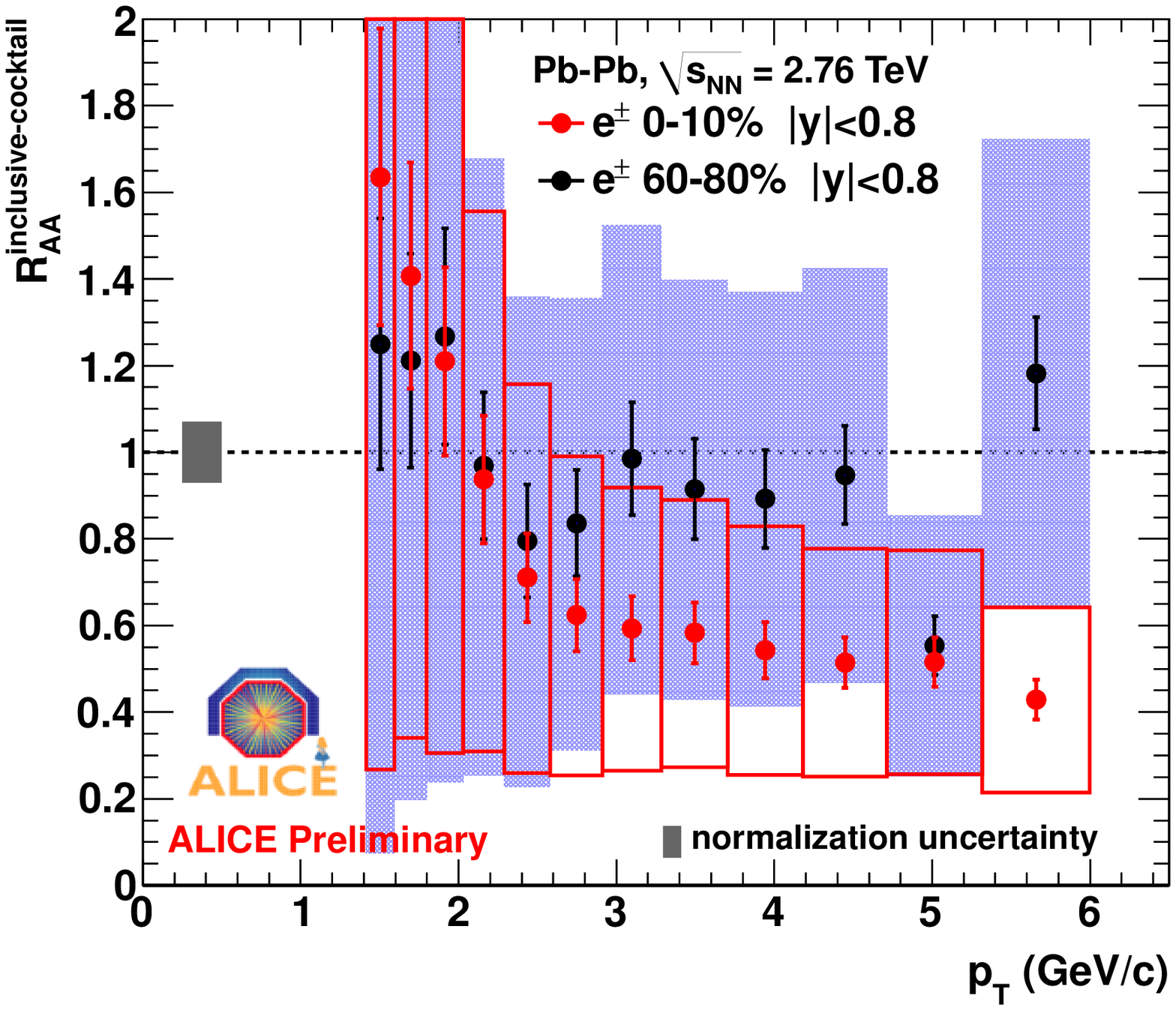}
    \includegraphics[height=0.23\textheight]{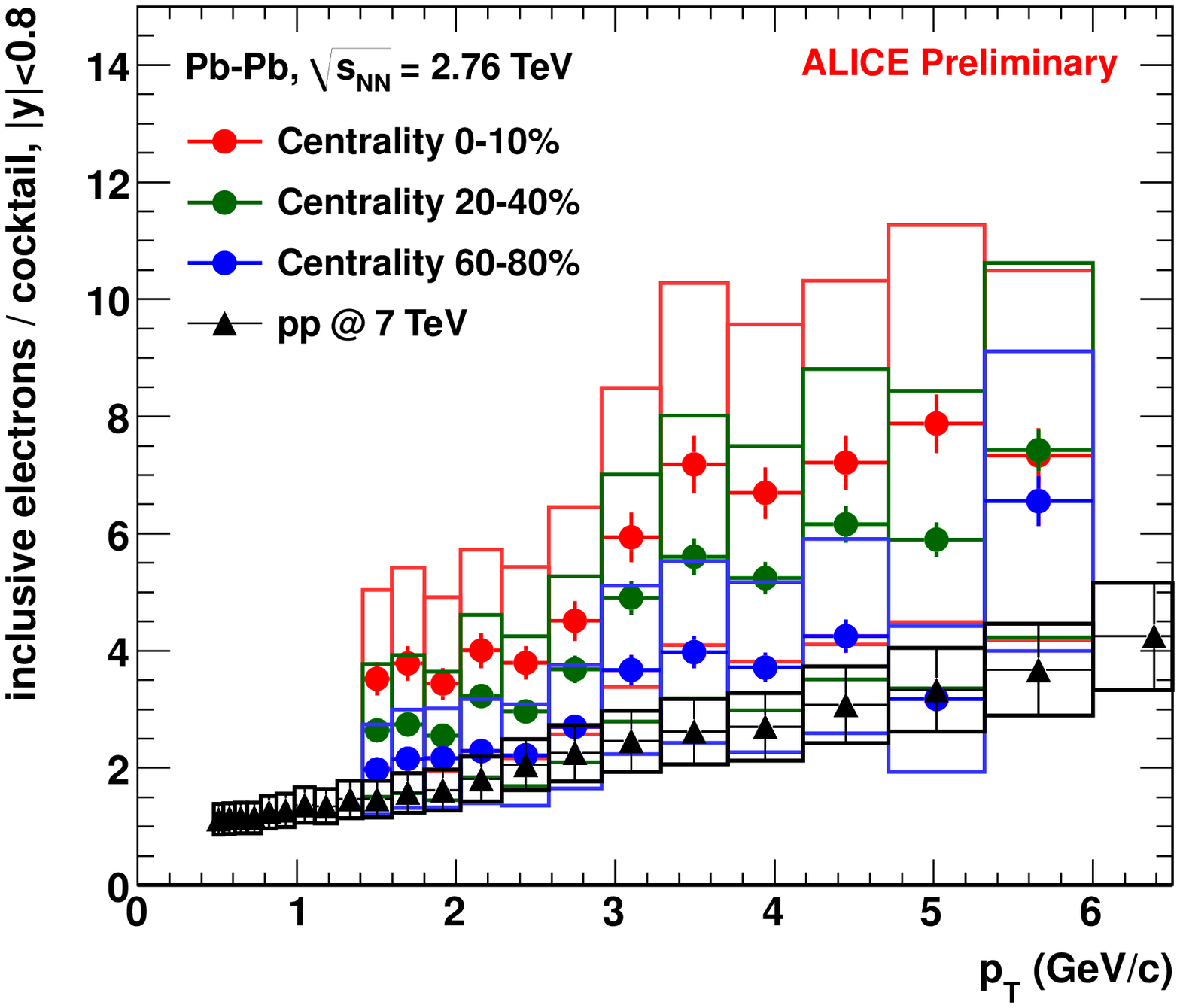}
    \caption[]{Left panel: $\RAA$ of electrons from heavy-flavour hadron decay in central and peripheral Pb--Pb collisions
      at $\sqrtsNN=2.76~\TeV$. Right panel: ratio between the inclusive electron spectrum and the spectrum of
      electrons not coming from heavy-flavour hadron decay, obtained with the cocktail technique (see text), in pp collisions
      and in Pb--Pb collisions at different centralities.
    }
    \label{fig:electronPbPb}    
  \end{center}
\end{figure}
\subsection{Reconstruction of D meson hadronic decays with the ALICE detector} 
\label{sec:Dreco}
ALICE has measured the production of $\Dzero$, $\Dplus$ ($\ctau\approx 123$ and $312~\mum$ respectively~\cite{pdg}) 
and $\Dstar$ (strong decay) in pp and Pb--Pb collisions at central rapidity ($|y|<0.5$) via the exclusive reconstruction 
of the decays $\DtoKpi$ (with branching ratio, BR=$3.89\pm 0.05\%$~\cite{pdg}) 
and $\DtoKpipi$ (BR=$9.4\pm 0.4\%$~\cite{pdg}) and $\DstartoDpi$ (BR=$67.7\pm0.5\%$~\cite{pdg}).
%
The analysis strategy for the extraction of the 
signals out of the large combinatorial background from uncorrelated tracks 
is based on the reconstruction 
and selection of secondary vertex topologies with significant separation (typically a few hundred micrometer) 
from the primary vertex~\cite{Dpp7TeVpaper}. 
The TPC and the ITS detectors provide a spatial resolution
on the track position in the vicinity of the primary vertex of the order of few tens of microns
at sufficiently high $\ptrans$~\cite{RossiVertex2010}.
Tracks with $|\eta|<0.8$ and displaced from the primary vertex 
are selected to reconstruct $\Dzero$ and $\Dplus$ meson candidates.
$\Dstar$ candidates are obtained by combining the $\Dzero$ candidates
with tracks with transverse momentum 
$\ptrans>0.08~\GeV/c$ in pp collisions and, in Pb--Pb collisions,
$\ptrans>0.2~\GeV/c$ in the centrality range 0--20\% and 
$\ptrans>0.1~\GeV/c$ in 20--80\%.
Good alignment between the reconstructed
meson momentum and the flight direction between the primary and secondary
vertex is also required.
The identification of the charged kaon in 
the TPC and TOF detectors provides additional background rejection. 
A particle identification strategy
that preserves most of the D meson signal was adopted.
Similar analyses were performed on $\pp$ and Pb--Pb data, with
a tighter selection in the latter case, dictated by the higher combinatorial
background.
To extract the signal, in the $\Dzero$ and $\Dplus$ cases a fit to the invariant mass distribution is performed,
while, for $\Dstar$ mesons, the mass difference $\Delta m=m_{\rm D^{*+}}-m_{\rm D^0}$ 
distribution is fitted. 
The `raw' signal is corrected for acceptance and
efficiency using Monte Carlo simulations based on Pythia (Perugia-0 tuning)~\cite{Pythia,Perugia0} 
and HIJING~\cite{Hijing} event generators. 
%
The contribution of D mesons from B decays was evaluated
relying on the FONLL prediction~\cite{fonll}, which 
describes well bottom production at the Tevatron~\cite{fonllBcdf} and at the 
LHC~\cite{lhcbBeauty,cmsJpsi}. 
The systematic error due to the FONLL theoretical uncertainty on 
beauty prediction was estimated from the spread 
of the results recovered 
using the minimum and maximum predictions for secondary D meson production. 
In the Pb--Pb case, 
the FONLL prediction of secondary D meson production in pp collisions
is multiplied by $\avNcoll \times \RAAB$, where
the hypothesis on the B meson $\RAA$ encodes all
potential nuclear and medium effects affecting B production. 
\section{Results from proton-proton collisions at $\sqrts=7~\TeV$}
Figure~\ref{fig:electronpp} shows on the left the $\ptrans$ differential cross section
of electrons from heavy-flavour hadron decay (black circles), as measured
on a minimum-bias sample of pp collisions at $\sqrts=7~\tev$ corresponding
to an integrated luminosity of ${\rm L_{int}}=2.6~{\rm nb^{-1}}$. 
The systematic uncertainty is about $20\%$ on the inclusive
electron spectrum, dominated by the electron identification uncertainty,
and about $10\%$ on the cocktail. At low $\ptrans$ electrons from charm hadron
constitute the main contribution, as clearly visible from
the comparison with the cross section of electron from D meson decay,
deduced from the D meson $\ptrans$-differential cross sections~\cite{Dpp7TeVpaper} by applying PYTHIA
decay kinematics and displayed in the same panel (pink square markers).
Electrons from b hadron decay can be selected by requiring
that the electron track is displaced with respect to the primary vertex of the collision.
The measured cross section, based on the analysis of a sub-sample with ${\rm L_{int}}=1.3~{\rm nb^{-1}}$ 
and reported on the right panel of Fig.~\ref{fig:electronpp},
is compatible with that obtained by subtracting the electron-from-D cross section from the 
heavy-flavour electron cross section. The measurements are well reproduced, 
within uncertainties, by FONLL predictions~\cite{fonll}, based on perturbative QCD calculations.
\begin{figure}[!t]
  \begin{center}
    \includegraphics[height=0.23\textheight]{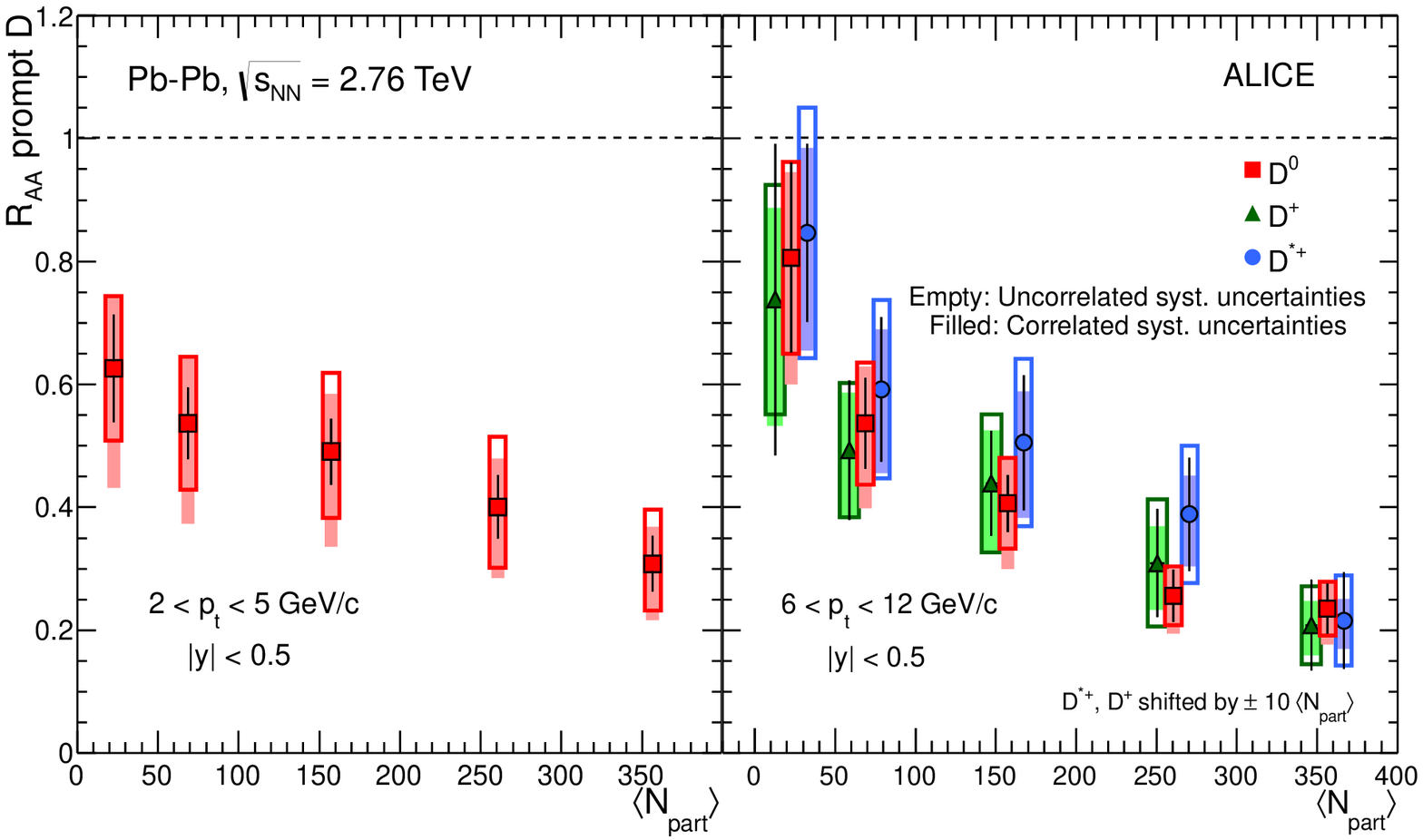}
    \includegraphics[height=0.24\textheight]{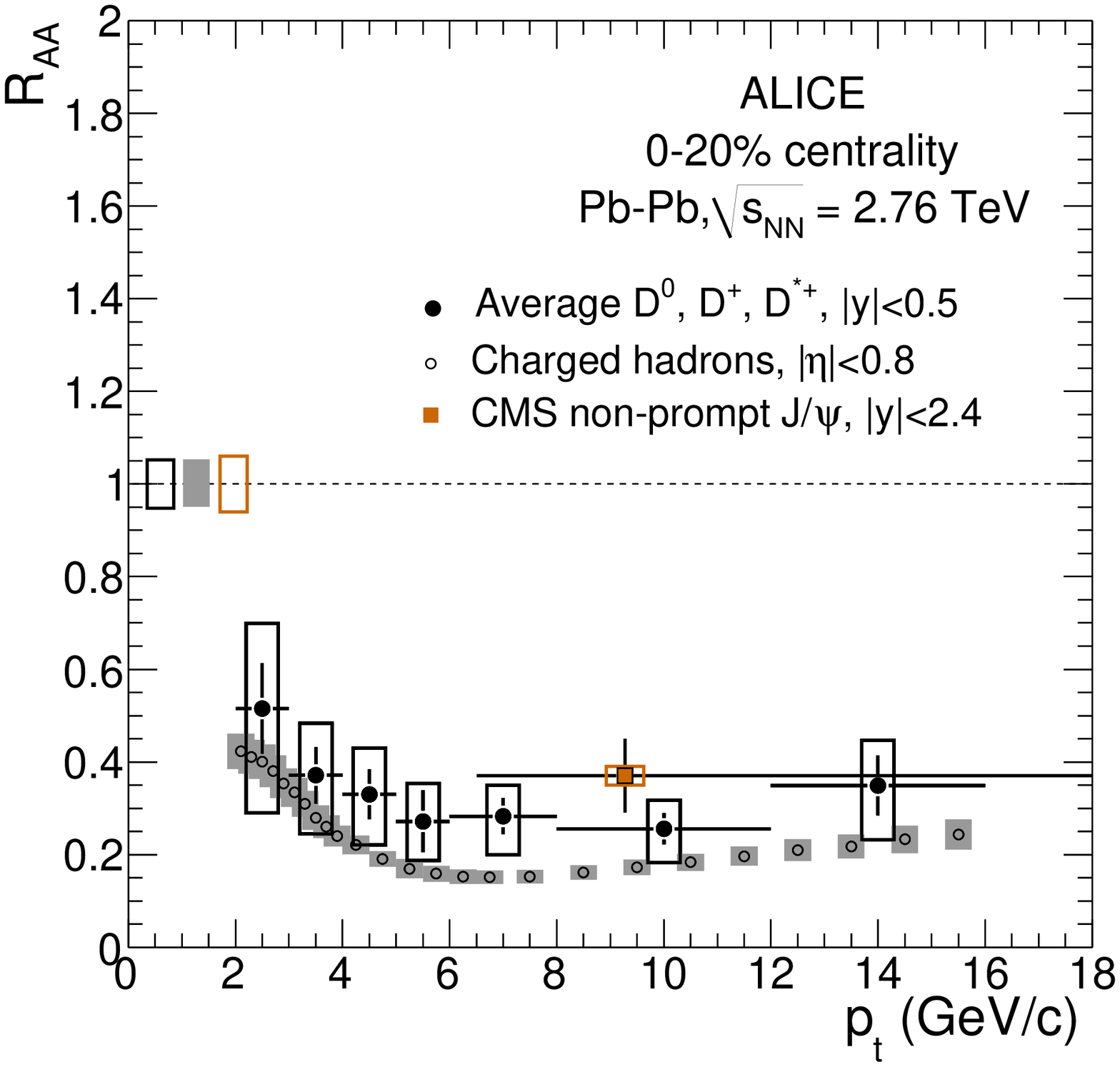}
    \caption[]{Left panel: $\Dzero$, $\Dplus$ and $\Dstar$ 
      nuclear modification factor measured in central  ($0-20\%$) 
      Pb--Pb collisions as a function
      of centrality for $2<\ptrans<5~\GeV/c$ (left) and $6<\ptrans<12~\GeV/c$ (right).
      The filled boxes represent the systematics uncertainties correlated
      among the different centrality intervals.
      Right panel: Average $\Dzero$, $\Dplus$ and $\Dstar$ 
      nuclear modification factor as a function of $\ptrans$ measured in central  ($0-20\%$) 
      Pb--Pb collisions, compared to the $\RAA$ of charged hadrons~\cite{HadronRAA} and 
      of non-prompt $\Jpsi$ from B decay~\cite{CMSjpsiPbPb}. 
      The boxes around $\RAA=1$ represent the uncertainty on the 
      normalization of the $\pp$ reference and on $\avTAA$. Both plots are taken from~\cite{RAADmeson}.
    }
    \label{fig:DmesonRAAVsPtVsNpart} 
  \end{center}
\end{figure}
%

As reported in~\cite{Dpp7TeVpaper}, ALICE has measured the  
$\ptrans$-differential production cross sections in pp collisions at $\sqrts=7~\TeV$
in the momentum range $1<\ptrans<16~\GeV/c$ for $\Dzero$
and in the range $1<\ptrans<24~\GeV/c$ for $\Dstar$ and $\Dplus$.
The measurements, based on the analysis of a minimum-bias sample
with ${\rm L_{int}}=5~{\rm nb^{-1}}$, are well described by predictions based on 
pQCD calculations as FONLL~\cite{fonll} and GM-VFNS~\cite{VFNSalice}.

To provide the reference cross section at $2.76~\tev$, needed to compute the $\RAA$,  
the measurements at $7~\tev$ are scaled by the ratio of FONLL predictions 
at $\sqrts=2.76$ and $7~\tev$~\cite{Note276}. 
In March 2011 a sample of $\approx 6.5\times 10^{7}$ events from
$\pp$ collisions at $\sqrts=2.76~\TeV$ was collected.  
The $\Dzero$ and $\Dplus$ signals were measured
in the range $2<\ptrans<8~\gev/c$.
For D mesons, while the accumulated statistics did not 
allow for determining a pp reference over the whole momentum range, 
it provided an important cross check of the theoretical scaling procedure 
in the momentum range where the data sets overlap. 
A similar measurement is being carried out for the electron decay channel
and, in this case, should provide a reduction of the systematic uncertainty
on the pp reference.
\section{Results from Pb--Pb collisions at $\sqrtsNN=2.76~\TeV$}
The data from Pb--Pb collisions at centre-of-mass energy $\sqrtsNN=2.76~\tev$
were collected with an interaction trigger based on the information 
of the SPD and the VZERO detector.
In total, $13\times 10^6$ Pb--Pb events with centrality in the range 
0--80\% were used in the analyses. The corresponding integrated luminosity is 
$L_{\rm int}=2.12\pm0.07~\mub^{-1}$. 
\subsection{Heavy-flavour electron nuclear modification factor}
Figure~\ref{fig:electronPbPb} shows on the left-hand panel the $\RAA$ of
cocktail-subtracted electrons in central ($0-10\%$) and
peripheral ($60-80\%$) Pb--Pb collisions. In peripheral collisions,
$\RAA$ is compatible with one. In central collisions,
a suppression is observed in the range $3.5<\ptrans<6~\GeV/c$ that,
due to the large uncertainties, can be estimated in a factor between 1.2 and 5.
On the right-hand panel, the ratio between the inclusive electron spectrum
and the background cocktail is reported for different centralities and 
for pp collisions. In pp collisions, the ratio approaches unity towards
low $\ptrans$, where the inclusive electron spectrum is 
dominated by electrons from background sources. In Pb-–Pb collisions, this ratio is 
larger, especially in the most central collisions,
causing the $\RAA$ rise towards low $\ptrans$. The effect, still under investigation,
suggests the presence of a further source of electrons: a possible candidate is  
thermal radiation, observed by the PHENIX experiment 
below $3~\GeV/c$ in Au--Au collisions at RHIC~\cite{PHENIXthermalGamma}.

\subsection{D meson nuclear modification factors}
The D meson production is suppressed by a factor 4-5 in central ($0-10\%$) events for $\ptrans\gsim 6~\gev/c$, 
as quantified by the nuclear modification factors shown 
in Fig.~\ref{fig:DmesonRAAVsPtVsNpart} as a function of the centrality,
expressed in terms of the average number of participant nucleons, $\Npart$, at
low (left, $\Dzero$ only) and high $\ptrans$ (middle).
At high $\ptrans$, $\RAA$ decreases from $\approx 0.81$ in peripheral ($\Npart=23$)
to $\sim 0.2$ in central ($\Npart=357$) events. 
The $\Dzero$,$\Dplus$ and $\Dstar$ $\RAA$
agree within errors. Their average, calculated using the statistical uncertainty as weight 
and taking into account the correlation among the systematic uncertainties of the different channels,
is shown on the right-hand panel of Fig.~\ref{fig:DmesonRAAVsPtVsNpart}
as a function of $\ptrans$. In the same panel the $\RAA$ of charged hadrons~\cite{HadronRAA}
and that, measured by CMS, of non-prompt $\Jpsi$ from b-hadron decay~\cite{CMSjpsiPbPb} are 
displayed.
In central collisions, $\RAA$ decreases with $\ptrans$ from $\sim 0.51$ at low $\ptrans$
to $\approx 0.27$ at high $\ptrans$. 
The average $\RAA$ is very close to the $\Dzero$ $\RAA$,
measured with a smaller statistical uncertainty ($15-20\%$) than for $\Dplus$ and $\Dstar$. 
The estimated total systematic uncertainty,
accounting for the uncertainties on the signal extraction procedure, on track reconstruction
efficiency, on the MC corrections for reconstruction
acceptance, cut and PID  selection, 
is of the order of $30~\%$ at intermediate $\ptrans$,
${}^{+35\%}_{-45\%}$ for $2<\ptrans<3~\gev/c$ and ${}^{+32\%}_{-40\%}$ for $12<\ptrans<16~\gev/c$.
The hypothesis on $\RAAB$, which is done to subtract the feed-down from B mesons as 
explained in section~\ref{sec:Dreco}, is varied in order to span the range $1/3<\RAAD/\RAAB<3$, with 
$\RAAD/\RAAB$ calculated a posteriori. 
The range of $\RAAD$ values obtained in each $\ptrans$ bin
is considered as the systematic error due to the $\RAAB$ assumption.
For the $\Dzero$, the maximum spread is ${}^{+14\%}_{-27\%}$ in the bin $12<\ptrans<16~\gev/c$.
%
\begin{figure}[!t]
  \begin{center}
    \includegraphics[height=0.23\textheight]{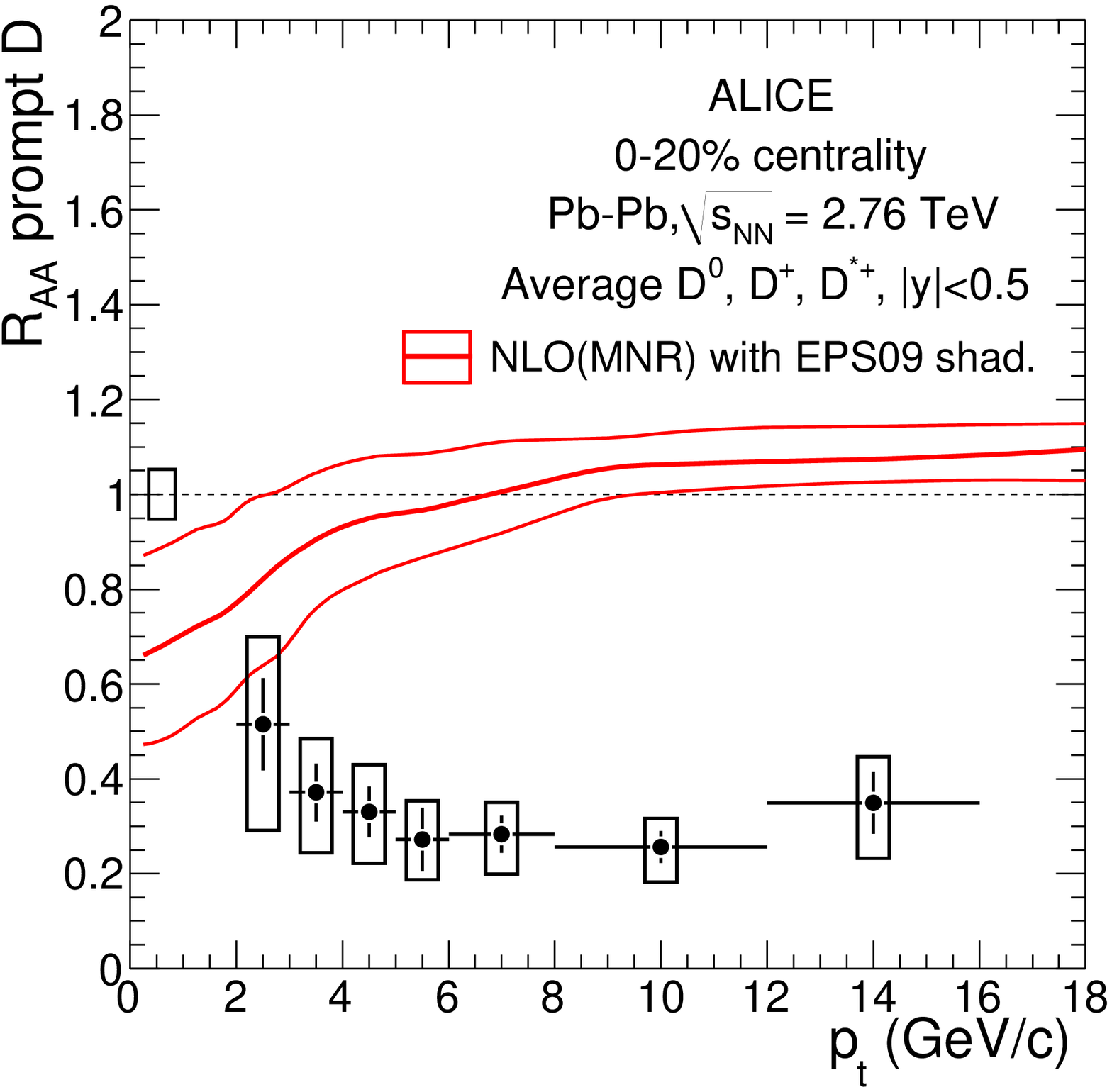}
    \includegraphics[height=0.23\textheight]{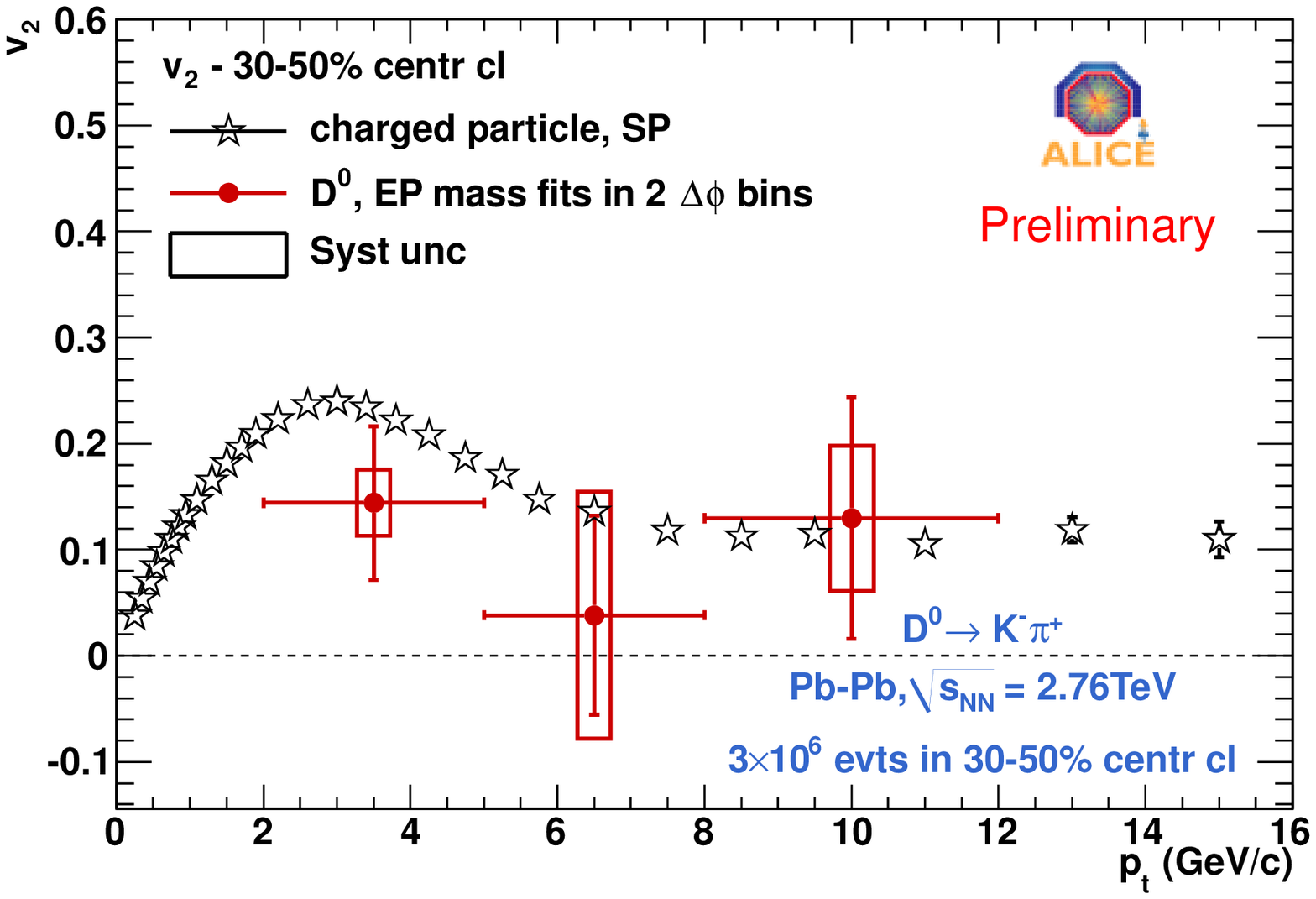}
    \caption[]{Left (from~\cite{RAADmeson}): Average $\RAA$ of D mesons in the 0--20\% centrality class compared to 
      the expectation from NLO pQCD~\cite{MNR} with nuclear shadowing~\cite{EPS09}.
      Right: $\Dzero$ $\vtwo$ as a function of $\ptrans$, obtained with the 2 $\Delta \phi$ method (see text).
      The vertical lines represent the statistical errors while the boxes the systematic uncertainties.
    }
    \label{fig:DraaVsShadowingDFlow}    
  \end{center}
\end{figure}
%
%

The D meson $\RAA$ is compatible, within errors, with the $\RAA$ of charged hadrons
and of non-prompt $\Jpsi$ from b-hadron decay. However,
considering that the systematic uncertainties of D mesons are not fully correlated
with $\ptrans$ there is an indication for $\RAA^{D}>\RAA^{\rm charged~hadrons}$. More precise and more
$\ptrans$-differential measurements of the non-prompt $\Jpsi$~$\RAA$ 
are necessary to conclude on the possible difference between charm and beauty
suppression.
In addition to final state effects, where parton energy loss would be predominant, 
also initial-state effects are expected to influence the measured $\RAA$.
In the kinematic range relevant for charm production 
at LHC energies, the main expected effect is nuclear shadowing, 
which reduces the parton distribution functions for partons with nucleon momentum fraction $x$ below $10^{-2}$. The effect 
of shadowing on the D meson $\RAA$ was estimated using the MNR 
next-to-leading order (NLO) 
perturbative QCD calculation~\cite{MNR} and the nuclear modification
of the parton distribution functions from the EPS09 parametrization~\cite{EPS09}. 
As shown in Fig.~\ref{fig:DraaVsShadowingDFlow} (left-hand panel) 
nuclear shadowing yields a relatively small 
effect for $\ptrans\gsim 5~\gev/c$ ($0.85<\RAAD<1.08$ at $\ptrans=5~\gev/c$). 
Therefore, the observed suppression can be considered 
as an evidence of in-medium charm quark 
energy loss. 
\section{$\Dzero$ meson elliptic flow}
The first $\Dzero$ $\vtwo$ measurement~\cite{BianchinFlow} was performed with event plane methods
and 2-particle cumulants~\cite{BorghiniFlow}. 
In the event plane methods, the correlation of $\Dzero$ azimuthal angle ($\phi$) 
to the reaction plane $\Psi_{\rm RP}$ is analyzed. 
The reaction plane is estimated via the event plane $\Psi_{2}$, by the so-called $Q_{2}$-vector, which is obtained
from a weighted sum of the azimuthal angles of all TPC tracks with $|\eta|<0.8$, satisfying 
quality requirements. 
Three different techniques were used to measure an ``observed" $\vtwo^{obs}$ which is
then corrected by the event-plane resolution, estimated from sub-events, to
measure $\vtwo$.  
In the first method, an in-plane and an out-of-plane region of $\Delta \phi=\phi-\Psi_{2}$ 
are defined and the signals $N_{IN}$ and $N_{N_{OUT}}$ are extracted by fitting 
the invariant mass distributions observed in the two regions. The $\vtwo^{obs}$ is recovered as 
\begin{equation}
\vtwo^{obs}=\frac{\pi}{4}\frac{N_{IN}-N_{OUT}}{N_{IN}+N_{OUT}}.
\end{equation}
In the second approach $\vtwo^{obs}$ is estimated as the average of the $\cos(2\Delta\phi)$ distribution
observed in the invariant mass $\Dzero$ signal region after the subtraction of the background distribution
obtained from the side bands. In the third method, 
the distribution of $\vtwo^{obs}(M)=\langle \cos(2\Delta\phi) \rangle (M)$ 
is fitted with a two component function, which includes a background term $\vtwo^{obs~back}(M)$
and a signal term $\vtwo^{obs~sign}$, weighted by the background and signal fractions, which are estimated
as a function of the invariant mass, from the invariant mass fit. A linear function is used for 
the $\vtwo^{obs~back}(M)$ component.
The $\vtwo$ measurement with the 2 $\Delta \phi$ technique is reported,
for the centrality range $30-50\%$, on the right-hand panel of Fig.~\ref{fig:DraaVsShadowingDFlow} as a function of $\ptrans$
and compared to the charged hadron $\vtwo$.  
The precision of the measurement is still limited by the available statistics~($3\times10^{6}$ events): 
within $1.8~\sigma$ 
a non-zero $\vtwo$ is measured for $\Dzero$ mesons in the range $2<\ptrans<5~\GeV/c$.
Within errors, compatible results are obtained with the techniques described above
and with the cumulant approach (see~\cite{BianchinFlow} for more details). The analysis 
of Pb--Pb data collected in 2011 will clarify whether charm $\vtwo$ is really
non-zero and how it does compare to lighter hadron elliptic flow.

\section{Summary}
ALICE has measured charm and beauty production in pp collisions at $7~\TeV$ and Pb--Pb collisions
at $\sqrtsNN=2.76~\TeV$ via the reconstruction of D meson hadronic decays and single electron
from heavy flavour hadron decays. Calculations based on perturbative QCD, like FONLL~\cite{fonll} and GM-VFNS~\cite{VFNSalice},
reproduce the $\ptrans$-differential cross section measured in pp collisions within the uncertainties.
In Pb--Pb collisions the observed suppression of the production of D mesons as well as of 
cocktail-subtracted electrons for $\ptrans>3.5~\GeV/c$ indicates strong coupling of heavy quarks to the medium created in central
heavy-ion collisions. The large uncertainties on the measurement
of $\Dzero$ elliptic flow does not allow to conclude whether charm is thermalized or not: the Pb--Pb data collected
in 2011 should provide the required statistics to answer to this question.



\begin{thebibliography}{10} 
\parskip 0pt
\begin{spacing}{0}
\bibitem{gyulassy} M. Gyulassy and M. Plumer, Phys. Lett. {\bf B243} (1990) 432.
\bibitem{bdmps} R. Baier, Y.~L. Dokshitzer, A.~H. Mueller, S. Peigne and D. Schiff, 
Nucl. Phys. {\bf B484} (1997) 265.
\bibitem{thoma}
M. H. Thoma and M. Gyulassy, Nucl. Phys. {\bf B351} (1991) 491.\\
E. Braaten and M. H. Thoma, Phys. Rev. {\bf D44} (1991) 1298; Phys. Rev. {\bf D44} (1991) 2625.
\bibitem{deadcone} Y.~L.~Dokshitzer, D.~E.~Kharzeev,  
Phys.\ Lett.\  {\bf B519 } (2001)  199. \\
N.~Armesto, C.~A.~Salgado and U.~A.~Wiedemann,  Phys.\ Rev.\  {\bf D69} (2004)  114003.\\
 M.~Djordjevic, M.~Gyulassy, Nucl. Phys. {\bf A733} (2004) 265.\\
B.-W. Zhang, E. Wang and X.-N. Wang, Phys. Rev. Lett. {\bf 93} (2004) 072301.\\
S.~Wicks, W.~Horowitz, M.~Djordjevic and M.~Gyulassy,
  Nucl.\ Phys.\  {\bf A783 } (2007)  493.
\bibitem{glauber}
R.~J.~Glauber in Lectures in Theoretical Physics, NY, 1959, Vol. 1, 315.\\
M.~Miller {\it et al.}, Ann.\ Rev.\ Nucl.\ Part.\ Sci.\ {\bf 57} (2007) 205.
\bibitem{Armesto:2005iq} N.~Armesto, A.~Dainese, C.A.~Salgado and U.A.~Wiedemann,
  Phys. Rev. {\bf D71} (2005) 054027.
\bibitem{Masciocchi2011} S. Masciocchi for the ALICE Coll., J. Phys. {\bf G38} (2011) 124069.
\bibitem{BailhacheSQM} R. Bailhache for the ALICE Coll., Acta Physica Polonica B {\it Proceedings supplement} Vol. 5, No. 2 (2012) 291.  
\bibitem{Dpp7TeVpaper} B. Abelev {\it et al.} [ALICE Coll.], JHEP {\bf 01} (2012) 128.
\bibitem{RAADmeson}B. Abelev {\it et al.} [ALICE Coll.]	arXiv:1203.2160v1, submitted to JHEP.
\bibitem{LoicPrimQCD} L. Manceau for the ALICE Coll., these proceedings.
\bibitem{ArmestoFlow} N. Armesto {\it et al.}, J. Phys. {\bf G35}, (2008) 054001.
\bibitem{MolnarFlow} D. Molnar, J. Phys. {\bf G31}, (2005) S421-S428.
\bibitem{PhenixFlow} A. Adare {\it et al.} [PHENIX Coll.], Phys. Rev. Lett. 98, (2007) 172301.
\bibitem{BianchinFlow} C. Bianchin for the ALICE Coll., Acta Physica Polonica B {\it Proceedings supplement} Vol. 5, No. 2 (2012) 335. 
\bibitem{aliceJINST}
  K.~Aamodt {\it et al.} [ALICE Coll.], JINST {\bf 3} (2008) S08002. 
\bibitem{AlbericaProcQM} A. Toia for the ALICE Coll., J. Phys. {\bf G38} (2011) 124007. 
\bibitem{pdg} K. Nakamura {\it et al.} (Particle Data Group), J. Phys. {\bf G37}, (2010) 075021.
\bibitem{RossiVertex2010}
A. Rossi {\it et al.} [ALICE Coll.], PoS(Vertex2010)017, arXiv:1101.3491.
\bibitem{Pythia}
  T. Sj\"{o}strand, S.~Mrenna, P.~Skands, JHEP \textbf{05} (2006) 026.
\bibitem{Perugia0} 
P. Z.~Skands, arXiv:0905.3418 [hep-ph] (2009).
\bibitem{Hijing} 
X.-N.~Wang and M.~Gyulassy, Phys. Rev. {\bf D 44} (1991) 3501. 
\bibitem{fonll}
M. Cacciari, M. Greco, P. Nason, JHEP {\bf 9805} (1998) 007.\\
M.~Cacciari, S.~Frixione and P.~Nason,
  JHEP {\bf 0103} (2001) 006.
\bibitem{fonllBcdf}
  M.~Cacciari {\it et al.}, JHEP {\bf 0407} (2004) 033; private communication.
\bibitem{lhcbBeauty}
  R. Aaij {\it et al.} [LHCb Coll.], Phys. Lett. B{\bf 694} (2010) 209-216.
\bibitem{cmsJpsi}
  V. Khachatryan {\it et al.} [CMS Coll.], arXiv:1011.4193.
\bibitem{VFNSalice} B.A.~Kniehl {\it et al.}, Phys. Rev. Lett. {\bf 96} (2006) 012001.\\
B.A. Kniehl, G. Kramer, I. Schienbein, and H. Spiesberger, in preparation.
\bibitem{Note276} R. Averbeck, N. Bastid, Z. Conesa del Valle, A. Dainese, X. Zhang, arXiv:1107.3243 (2011).
\bibitem{HadronRAA} ALICE Collaboration, Centrality dependence of charged-hadron transverse momentum distributions in Pb--Pb collisions at $\sqrtsNN=2.76~\tev$, article in preparation. 
\bibitem{CMSjpsiPbPb}CMS Collaboration, arXiv:1201.5069 [nucl-ex] (2012).
\bibitem{PHENIXthermalGamma}A. Adate {\it et al.} [PHENIX Coll.], Phys. Rev. Lett. 104, (2010) 132301.
\bibitem{BorghiniFlow} N. Borghini, P. M. Dinh, J. Y. Ollitrault, Phys. Rev. {\bf C64} (2001) 054901.
\bibitem{RaimondFlow} A. Bilandzic, R. Snellings, and S. Voloshin, Phys. Rev. {\bf C83} (2001) 044913.
\bibitem{MNR} M.~Mangano, P.~Nason and G.~Ridolfi, Nucl.~Phys.~\textbf{B373} 
(1992) 295.
\bibitem{EPS09} K.~J.~Eskola, H.~Paukkunen and C.~A.~Salgado, JHEP {\bf 04} (2009) 065.
\end{spacing}
\end{thebibliography}
\end{document}